# Derivation of the paraxial form of the angular momentum of the electromagnetic field from the general form

## A. M. STEWART

Department of Theoretical Physics,
Research School of Physical Sciences and Engineering,
The Australian National University,
Canberra, ACT 0200, Australia.     4 September 2006

Keywords: angular momentum, electromagnetism, paraxial

**Abstract.** It is shown how the standard forms for the spin and orbital components of the angular momentum of a paraxial wave of electromagnetic radiation are obtained from the general expressions for the angular momentum that have been derived recently. This result will enable the general expressions for angular momentum to be applied with confidence to the many configurations of electromagnetic fields that are more complicated than plane or paraxial waves.

## 1. Introduction

The general expression for the angular momentum $J(t)$ of the classical electromagnetic field is given in terms of the electric $E(r,t)$ and magnetic $B(\mathbf{r},t)$ Maxwell fields by [1]

$$J(t) = \frac{1}{4\pi c} \int d^3 r\, \mathbf{r} \times [E(\mathbf{r},t) \times B(\mathbf{r},t)] \tag{1}$$

(Gaussian units, bold font denotes a three-vector). It has been shown elsewhere [2-5] that by expressing the electric field vector as the sum of its longitudinal and transverse components by means of the Helmholtz theorem [6]

$$E(\mathbf{r},t) = \nabla \times \int_V dV' \frac{\nabla' \times E(\mathbf{r}',t)}{4\pi |\mathbf{r}-\mathbf{r}'|} - \nabla \int_V dV' \frac{\nabla'\cdot E(\mathbf{r}',t)}{4\pi |\mathbf{r}-\mathbf{r}'|} , \tag{2}$$

where $\nabla'$ is the gradient operator with respect to $\mathbf{r}'$, the angular momentum of the electromagnetic field away from the influence of charge [5] is given by the sum of a spin term

$$J_s = \frac{1}{(4\pi c)^2} \int_V d^3 r \int_V d^3 r' \frac{B(\mathbf{r},t)}{|\mathbf{r}-\mathbf{r}'|} \times \frac{\partial B(\mathbf{r}',t)}{\partial t} \tag{3}$$

and an orbital term

$$J_o = \frac{1}{(4\pi c)^2} \int_V d^3 r \int_V d^3 r' [B(\mathbf{r},t) \cdot \frac{\partial B(\mathbf{r}',t)}{\partial t}] \frac{\mathbf{r} \times \mathbf{r}'}{|\mathbf{r}-\mathbf{r}'|^3} . \tag{4}$$

There is also a surface term whose existence is important for understanding the apparently paradoxical properties of the angular momentum of plane waves [4] but which is not needed for the discussion in the present paper.

In the past decade or more [7-12] there has been much work on the angular momentum properties of paraxial electromagnetic waves, which describe laser light with great accuracy. Very simple expressions have been derived for the spin and orbital components of the angular momentum of paraxial waves and these have been confirmed experimentally. The issue that is examined in the present paper is whether these simple expressions can be derived from the general expressions (3) and (4) above. It is shown that they can. This result gives confidence that (3) and (4) will be able to be applied to configurations of fields that are more complicated than plane or paraxial waves.

## 2. The paraxial approximation

In this section we summarize results for angular momentum that have been reported to have been obtained in the paraxial approximation for a wave of cylindrical cross-section with radius $R$ travelling in the $z$ direction [7, 9]. It is assumed that the electromagnetic vector potential is of the form $\mathbf{A} = \mathbf{S}f$ with polarization vector $\mathbf{S} = \alpha\hat{\mathbf{x}} + \beta\hat{\mathbf{y}}$, $\alpha$ and $\beta$ being complex numbers with $\alpha\alpha^* + \beta\beta^* = 1$, the star denoting the complex conjugate, and the complex scalar $f$ given by

$$f = bv(r)e^{-il\phi}e^{i(kz-\omega t)}, \qquad (5)$$

where the cylindrical coordinates are $(r,\phi,z)$. The quantity $b$ is a real amplitude and $v(r)$, which is a real function of $r$ and falls off above a radius $R$, is normalized to $\int_0^\infty v(r)^2 r \, dr = 1$. This is the form of paraxial waves that possesses orbital angular momentum about the $z$-axis. The wavelength is small compared to the transverse diameter $2R$ of the wave so that $kR \gg 1$. For a red laser of beam diameter 2 mm the value of $kR$ is around 10,000. The large value of this factor will be important in later stages of the calculation.

The magnetic field is given by $\mathbf{B} = -\mathbf{S} \times \nabla f$ and the electric field is obtained from the Lorenz gauge condition [9]. Using (1), the density of angular momentum in the $z$ direction is reported to come to [9]

$$j^z = \varepsilon_0 l\omega b^2 v^2 - \frac{\varepsilon_0}{2}\sigma\omega b^2 r \frac{\partial v^2}{\partial r}, \qquad (6)$$

where the polarization parameter is $\sigma = i(\alpha^*\beta - \alpha\beta^*)$. To obtain the angular momentum in the $z$ direction $J^z$ contained in length d$z$ this expression is integrated from 0 to $2\pi$ over $\phi$ and from 0 to infinity over $r$ weighted by $r$. The first term gives the orbital contribution

$$J_o^z = 2\pi\varepsilon_0 l\omega b^2 \, dz \, . \tag{7}$$

The second term of (6), after a partial integration, gives the spin contribution

$$J_s^z = 2\pi\varepsilon_0 \sigma\omega b^2 \, dz \, , \tag{8}$$

the total angular momentum coming to $J^z = 2\pi\varepsilon_0(\sigma + l)\omega b^2 \, dz$. The purpose of the present paper is to show how equations (7) and (8) are obtained from (3) and (4) using the form for the fields of a paraxial wave given by (5).

### 3. The spin component of the angular momentum of a paraxial wave

From (3), and taking account of the complex nature of the electromagnetic fields of a paraxial wave given by (5),

$$\boldsymbol{J}_s = \mathrm{Re}\frac{i\omega}{(4\pi c)^2} \int_V d^3r \int_V d^3r' \frac{[\boldsymbol{S}\times\nabla f(\boldsymbol{r})]\times[\boldsymbol{S}^*\times\nabla' f^*(\boldsymbol{r}')]}{|\boldsymbol{r}-\boldsymbol{r}'|} . \tag{9}$$

The double vector cross-product may be worked out to be, in cylindrical coordinates $\boldsymbol{r} = (r,\phi,z)$ and $\boldsymbol{r}' = (r',\phi',z')$,

$$[\boldsymbol{S}\times\nabla f(\boldsymbol{r})]\times[\boldsymbol{S}^*\times\nabla' f^*(\boldsymbol{r}')] = i\hat{\boldsymbol{z}}\sigma \frac{\partial f}{\partial z}\frac{\partial f^*}{\partial z'} . \tag{10}$$

From (5), and using the property of the paraxial approximation that the derivatives in the $z$ direction predominate over those in the transverse directions, we find

$$[\boldsymbol{S}\times\nabla f(\boldsymbol{r})]\times[\boldsymbol{S}^*\times\nabla' f^*(\boldsymbol{r}')] = i\hat{\boldsymbol{z}}\sigma k^2 b^2 v(r)v(r')e^{ik(z-z')}e^{-il(\phi-\phi')} . \tag{11}$$

The quantity $|\boldsymbol{r}-\boldsymbol{r}'|$ is given in cylindrical coordinates by

$$|\boldsymbol{r}-\boldsymbol{r}'| = \sqrt{(z-z')^2 + r^2 + r'^2 - 2rr'\cos(\phi-\phi')} \, , \tag{12}$$

giving for the spin component of the angular momentum in the $z$ direction

$$J_s^z = -\mathrm{Re}\frac{\omega\sigma k^2 b^2}{(4\pi c)^2} \int_V d^3r\, v(r) \int_V d^3r' \frac{v(r')e^{-il(\phi-\phi')}e^{ik(z-z')}}{\sqrt{(z-z')^2 + \beta^2}} . \tag{13}$$

The quantity $\beta$ is given by $\beta = r\sqrt{(1-a)^2 + 4a\sin^2[(\phi-\phi')/2]}$ with $a = r'/r$ ($0 < a < \infty$).

First we do the integral over $z'$ in (13) with volume element $d^3r' = dz'd\phi'r'dr'$ by making the substitution $x = z - z'$. The integral of the sine term in the exponential of $ik(z - z')$ vanishes by symmetry. The cosine term leads to the integral

$$\int dx \frac{\cos(kx)}{\sqrt{x^2+\beta^2}} = 2K_0(k\beta) \quad . \tag{14}$$

$K_0$ is the Bessel function of imaginary argument [13] and index 0. Because the Bessel function has a logarithmic divergence at zero argument and decays monotonically and faster than exponentially for large argument [13] it follows that contributions to the integral in (13) will occur only around $a = 1$ (or $r' = r$) and $\phi = \phi'$ so these conditions may be imposed on the slowly varying factors. In the $\exp[-il(\phi - \phi')]$ factor in (13) the integral of the sine part vanishes by symmetry and the cosine part becomes unity, so the integral become real.

$$J_s^z = \frac{2\sigma\omega k^2 b^2}{(4\pi c)^2} \int_V d^3 r\, v(r)^2 r^2 \int_0^\pi a\, da \int_{-\pi}^{\pi} d\phi'\, K_0[kr\sqrt{(1-a)^2 + 4\sin^2[(\phi-\phi')/2]}] \quad . \tag{15}$$

Now, setting $m = 1 - a$ and $n = \phi - \phi'$, we get

$$J_s^z = \frac{2\sigma\omega k^2 b^2}{(4\pi c)^2} \int_V d^3 r\, v(r)^2 r^2 \int_0^\pi dm \int_{-\pi}^{\pi} dn\, K_0[kr\sqrt{m^2 + 4\sin^2(n/2)}] \quad .\tag{16}$$

The limits of integration may be extended to +/- infinity and the sine may be replaced with its argument to give

$$J_s^z = \frac{\sigma\omega k^2 b^2}{(4\pi c)^2} \int_V d^3 r\, v(r)^2 r^2 \int_{-\infty}^{\infty} dm \int_{-\infty}^{\infty} dn\, K_0(kr\sqrt{m^2+n^2}) . \tag{17}$$

Next, by introducing the radial cylindrical coordinate $\rho$ given by $\rho = \sqrt{m^2+n^2}$, the double integral over $m$ and $n$ becomes the single integral over $\rho$

$$J_s^z = \frac{\sigma\omega k^2 b^2}{(4\pi c)^2} \int_V d^3 r\, v(r)^2 r^2 \int_0^\infty d\rho\, 2\pi\rho K_0(kr\rho) . \tag{18}$$

The integral over $\rho$ comes to $2\pi/(kr)^2$ [13], giving

$$J_s^z = \frac{\sigma\omega b^2}{8\pi c^2} \int_V d^3 r\, v(r)^2 \quad . \tag{19}$$

The radial and angular integrations of $d^3 r$ may then be performed to give

$$J_s^z = \frac{\sigma\omega b^2}{4c^2} \int dz \quad . \tag{20}$$

**4. The orbital component of the angular momentum of a paraxial wave**
The term in (4) involving the polarization vector comes to

$$[\mathbf{S} \times f(\mathbf{r})] \cdot [\mathbf{S}^* \times {'f}^*(\mathbf{r}')] = k^2 b^2 v(r) v(r') e^{ik(z-z')} e^{-il(\phi-\phi')} \quad . \quad (21)$$

and from $\mathbf{r} \times \mathbf{r}'|^z = rr'\sin(\phi'-\phi)$ we get, with $a = r'/r$,

$$J_0^z = i\frac{\omega k^2 b^2}{(4\pi c)^2} \int_V d^3r\, v(r) r^2 \int_V d^3r' \frac{v(r') a \sin(\phi'-\phi) e^{-il(\phi-\phi')} e^{ik(z-z')}}{[(z-z')^2 + \beta^2]^{3/2}} \quad . \quad (22)$$

Again we do the integral over $z'$, using the substitution $x = z - z'$, to get

$$\int_{-\infty}^{\infty} dx\, \frac{\cos(kx)}{(x^2 + \beta^2)^{3/2}} = \frac{2k}{\beta} K_1(k\beta) \,, \quad (23)$$

where $K_1$ is the Bessel function of imaginary argument [13] and index 1. The integral (23) now becomes

$$J_0^z = \operatorname{Re} \frac{i2\omega k^3 b^2}{(4\pi c)^2} \int_V d^3r\, v(r) r \int_0^\infty dr'\, r' v(r') \int_{-\pi}^{\pi} d\phi' \frac{a e^{-il(\phi-\phi')} \sin(\phi-\phi') K_1(k\beta)}{\sqrt{(1-a)^2 + 4a\sin^2[(\phi-\phi')/2]}} \quad . \quad (24)$$

The integral of the cosine part of the $\exp[-il(\phi - \phi')]$ factor in (24) vanishes by symmetry and the sine part may be replaced by its argument.

The Bessel function $K_1(x)$ diverges as $1/x$ as $x \to 0$ and decays faster than exponentially at large $x$. For $kr \gg 1$ it is clear that the main contributions to the integral will come from around $a = 1$ and $\phi = \phi'$; the denominator also produces a divergence at these values of the parameters. These values may therefore be put in the factors outside the integral. Making the substitutions $m = 1 - a$, $n = \phi - \phi'$ we get

$$J_0^z = \frac{l\omega k^3 b^2}{(4\pi c)^2} \int_V d^3r\, v(r)^2 r^3 \int_{-\infty}^{\infty} dm \int_{-\infty}^{\infty} dn\, n^2 \frac{K_1(kr\sqrt{m^2+n^2})}{\sqrt{m^2+n^2}} \quad . \quad (25)$$

If, in the $m$ and $n$ integrals in the above equation, we interchange $m$ and $n$, we get the same result except with an $m^2$ instead of an $n^2$. Adding the two together, dividing by two, and again introducing the radial cylindrical coordinate $\rho$ given by $\rho = \sqrt{m^2+n^2}$, we get

$$J_0^z = \frac{l\omega k^3 b^2}{2(4\pi c)^2} \int_V d^3r\, v(r)^2 r^3 \int_0^\infty d\rho\, 2\pi \rho^2 K_1(kr\rho) \quad (26)$$

The second integral in (26) comes to $4\pi/(kr)^3$ [13] giving the result

$$J_0^z = \frac{l\omega b^2}{4c^2} \int dz \quad . \quad (27)$$

## 5. Discussion
Starting from the expressions (3) and (4) for the spin and orbital components of the angular momentum of the general electromagnetic field we have obtained the results (20) and (27) for the angular momentum of a paraxial wave described by (5), the different constants of proportionality to those of (7) and (8) being a consequence of the units used. The general expressions given at the beginning of this paper give the correct result for plane waves [2] and have enabled the paradox associated with the boundaries of a plane wave to be resolved [4]. These results, together with those of this paper, which confirm the validity of the paraxial results (7) and (8), will enable the general expressions for angular momentum (3) and (4) to be applied with confidence to the many configurations of electromagnetic fields that are more complicated than plane or paraxial waves.